\documentclass[twocolumn]{aastex63}

\usepackage{graphicx}
\usepackage{bm}
\usepackage{amsmath}
\usepackage{amssymb}
\usepackage{hyperref}

\received{ }
\revised{ }
\accepted{ }

\submitjournal{ApJ}

\shorttitle{Formation of complex PAHs}
\shortauthors{Hanine et al.}

\begin{document}

\title{Formation of interstellar complex polycyclic aromatic hydrocarbons: Insights from molecular dynamics simulations of dehydrogenated benzene}

\correspondingauthor{Zhao Wang}
\email{zw@gxu.edu.cn}

\author{Meriem Hanine}
\affil{Laboratory for Relativistic Astrophysics, Department of Physics, Guangxi University, 530004 Nanning, China}

\author{Zhisen Meng}
\affil{Laboratory for Relativistic Astrophysics, Department of Physics, Guangxi University, 530004 Nanning, China}

\author{Shiru Lu}
\affil{Laboratory for Relativistic Astrophysics, Department of Physics, Guangxi University, 530004 Nanning, China}

\author{Peng Xie}
\affil{School of Chemistry and Chemical Engineering, Guangxi University, 530004 Nanning, China}

\author{Sylvain Picaud}
\affil{Observatoire de Besan\c{c}on, Institut UTINAM, CNRS UMR 6213, UBFC, 25030 Besan\c{c}on, France}
	
\author{Michel Devel}
\affil{FEMTO-ST institute, CNRS, ENSMM, 15B avenue des Montboucons, 25030 Besan\c{c}on, France}

\author{Zhao Wang}
\affil{Laboratory for Relativistic Astrophysics, Department of Physics, Guangxi University, 530004 Nanning, China}

\begin{abstract}

Small organic molecules are thought to provide building blocks for the formation of complex interstellar polycyclic aromatic hydrocarbons (PAHs). However, the underlying chemical mechanisms remain unclear, particularly concerning the role of interstellar dust. Using molecular dynamics, we simulate the chemical reaction between dehydrogenated benzene molecules in the gas phase or on the surface of an onion-like carbon nanoparticle (NP). The reaction leads to the formation of PAHs of complex structures. The size of the formed molecules is found to roughly increase with increasing temperature up to 800 K, and to be correlated with the level of dehydrogenation. Morphology analysis features the formation of large rings that contain up to 32 carbon atom at high temperature. Density functional theory (DFT) calculations are performed to search the fundamental energetic reaction pathways. The DFT results quantitatively confirm the correlation between the reactivity and the dehydrogenation level, and the formation of stable C-8 rings. Moreover, the nanostructures formed on the NP surface point to a possible layer-by-layer formation mechanism for interstellar fullerene and carbon onions.

\end{abstract}


\section{Introduction} \label{sec:intro}

Polycyclic aromatic hydrocarbons (PAHs) are among the most studied molecules in the fields of chemistry, astronomy, biology and environmental science \citep{Keyte2013}. PAHs are found to be highly abundant in the interstellar medium (ISM), and to play important roles in its evolution \citep{Tielens2008}. Small interstellar organic molecules have been hypothesized to be involved in the formation of complex PAHs, which in turn are believed to be building blocks of complex organic molecules \citep{Ehrenfreund2006,Rapacioli2006,Ehrenfreund2000}. The building sequence could be crucial in explanations of ISM evolution and abiogenesis \citep{Wakelam2008,Galliano2008,Kim2012,Sandstrom2012,Puzzarini2017}. However, the underlying physicochemical mechanisms of the formation process of large and complex PAHs remain unclear, particularly regarding the key stage of chemical reactions between small organic compounds involving interstellar dust.

Recent space- and ground-based observations via infrared radiation (IR) spectroscopy have revealed the abundance of diverse carbon nanostructures in the interstellar dust, including fullerene, diamond, graphite, soot, and so forth \citep{Li2019,Bernal2019,Zhang2011,Boi2017}. They were reported to co-exist with PAHs in various regions of ISM and likely to influence the chemical evolution of PAHs \citep{Cami2010,Garcia-Hernandez2010,Chhowalla2003}. In one of our recent works, we have studied the interaction between adsorbed organic molecules on a fullerene-like carbon onion by means of classical molecular dynamics (MD) simulations \citep{Qi2018}. Calculations of molecular binding energy have suggested that carbon nanostructures could provide substrates for forming molecular layers and aggregations. However, the chemical reaction remains unexplored since we only studied fully saturated molecules. 

From a chemical perspective, high energy barrier is expected for the interaction between fully hydrogenated PAHs \citep{Chen2018}. Photofragmentation involving the dehydrogenation in ISM regions with strong UV irradiation is thus considered as the main mechanism leading to the formation of new PAH species. For instance, \cite{Castellanos2018} have demonstrated that the balance between H-loss due to UV irradiation and reactions with atomic H shifts towards gaseous dehydrogenated PAHs. The existence of dehydrogenated PAHs in ISM has also been suggested by Mackie et al. based on numerical analysis of Spitzer IR emission spectra \citep{Mackie2015}. A recent laboratory study has reported the formation of large PAHs following photodissociation and photodehydrogenation processes of PAH clusters \citep{Zhen2018}. In light of these evidences, we study chemical reaction between unsaturated organic molecules using MD based on a reactive force field, in order to gain insights into the formation of large and complex structures of PAHs. These simulations are performed for dehydrogenated benzene either in the gas phase or on a carbon onion-like nanoparticle (NP) \citep{Boi2017,Krasnokutski2017}. Density functional theory (DFT) calculations were also performed to identify fundamental reaction pathways. We explore the complex structures of the formed PAHs by means of morphology analyses.

\section{Methods} \label{sec:method}

\begin{figure}[htp]
\centerline{\includegraphics[width=9cm]{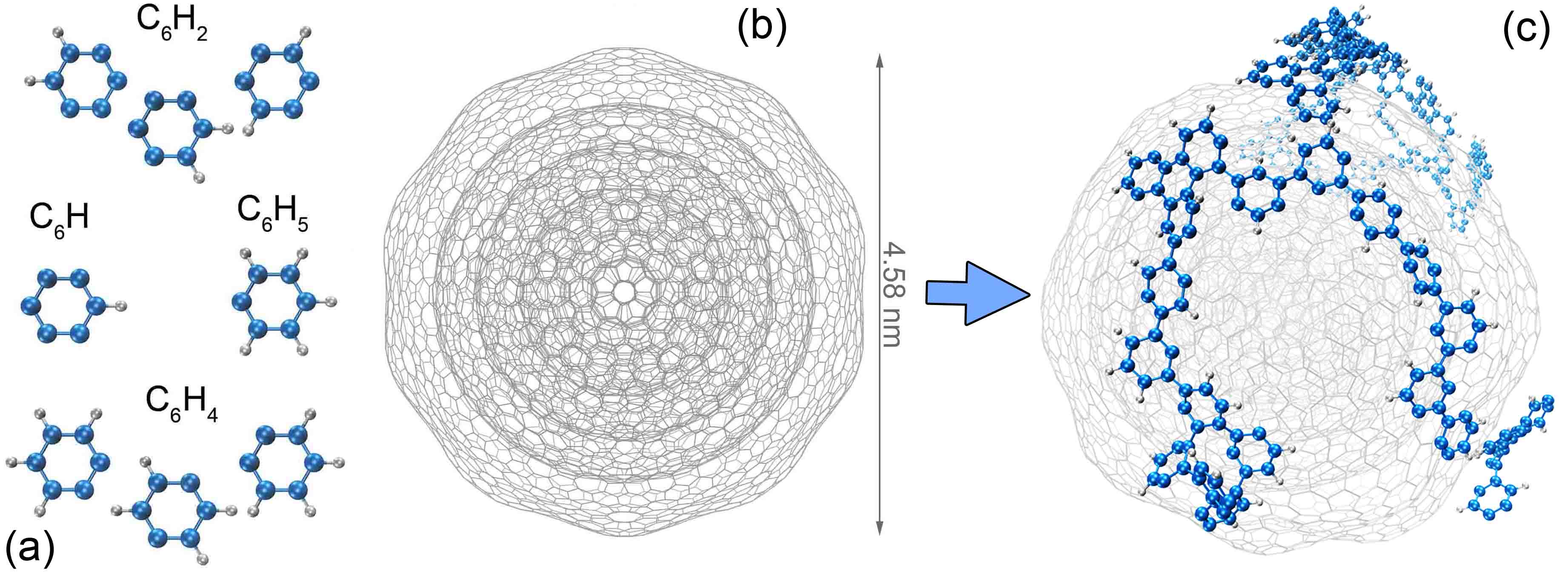}}
\caption{\label{F1}
(a) Ball-and-stick model of the dehydrogenated organic compounds studied in this work. Carbon atoms are depicted in blue, hydrogens in gray. (b) Line model of the onion-like fullerene structure of the carbon NP (diameter $=4.58$ nm). (c) Snapshot of the simulation cell that contains PAH molecules formed on the NP from C$_{6}$H$_{2}$ at $300$K.}
\end{figure}

It is known that dehydrogenation can occur in ISM environments through irradiation with, for instance, cosmic rays or ultraviolet (UV) photons \citep{Piani2017}, as evidenced by the discovery of unsaturated molecules in ISM \citep{Herbst2009,Mackie2015}. Even slight dehydrogenation could make molecules highly reactive \citep{Chen2018,Petrignani2016,Krasnokutski2017}. To simulate the formation of PAHs, we here take the smallest aromatic unit, benzene, with different levels of dehydrogenation (C$_{6}$H$_{5}$,C$_{6}$H$_{4}$,C$_{6}$H$_{2}$ and C$_{6}$H$_{1}$) as the samples shown in Figure \ref{F1} (a). The carbon NP is modeled by a bucky-onion containing several concentric fullerene layers as shown in Figure \ref{F1} (b), in order to mimic the onion-like nanostructures observed in transmission electron microscopy experiments \citep{Krasnokutski2017,Rotundi1998,Jager2006,Jager2009}. This model is in general consistent with the previously-reported low H/C ratio and high aromaticity of the galactic hydrocarbon dust \citep{Chiar2013}.

In our MD model, the interatomic interaction is described by a many-body force field, the adaptive interatomic reactive empirical bond order (AIREBO) model, which enables the formation or the breaking of chemical bonds using a spline-smoothed transition from long-range van der Waals (vdW) interaction to covalent bonding, or viceversa. Details of this force field, including its formulation, parameter values and benchmark calculations are provided in \cite{Stuart2000}. This force field has been widely used for simulating $sp^{2}$ hydrocarbons, and has shown reasonable accuracy in modeling chemical reaction \citep{Liu2011} and physisorption \citep{Wang2019,Wang2020a} of organic molecules.

A periodic simulation box (about $10 \times 10 \times 10$ nm$^{3}$ in size) is prepared, in which $60$ organic molecules and a carbon NP are placed at random sites in vicinity. The parallel computing package LAMMPS was used for the MD simulations {\citep{Plimpton1995}}. After initial velocities are randomly assigned to each atoms, the NP is brought to thermal equilibrium by a Nos\'{e}-Hoover thermostat. Note that adsorption of most of the organic molecules occurs on the NP surface during equilibration, where these molecules can start to form new species through chemical reactions. The cases of the gas phase (without NP) are also simulated. The molecules are considered to be dehydrogenated before coming to the NP surface. It is assumed that, in the low-temperature regions of ISM, the molecules with low-diffusivity may undergo photodissociation due to radiations in the gas phase before they collide with other molecules or dust. Since the temperature of ISM can vary a lot in different environments \citep{Ferriere2001,Zheng2008,Pino2019,Fu2012}, we simulate the reaction between the organic molecule at different temperatures, including $10$, $100$, $200$, $300$, $600$, $800$ and $1000$ K, in order to gain insight on the temperature dependence. With a time step of $0.5$ fs, the duration of every simulation is set to be $4.0$ ns (i.e. 8,000,000 steps). Despite of the radical density will be much lower in space \citep{Elmegreen2007}, the molecular density is here set to be high enough in order to speed up the reaction simulation for the sake of calculability, by considering that many collisions will occur over time.

\section{Results and discussions} \label{sec:RandD}
\subsection{Statistical distribution}

PAH molecules of different topology are observed to form as shown in Figure \ref{F1} (c) for example. The atomistic configurations of the formed molecules are recorded at different instance during the simulations. It is observed that small molecules spontaneously bind to each other to form small PAHs, which further cluster to form large and complex structures. In contrast to C$_{6}$H$_{2}$ and C$_{6}$H, C$_{6}$H$_{5}$ and C$_{6}$H$_{4}$ only produces small PAHs such as diphenyls. We therefore focus on the complex structures of PAHs formed from C$_{6}$H and C$_{6}$H$_{2}$ in the following discussion. Note that a detailed reaction sequence is illustrated via an animation provided in Wang (2020) which is available at 10.5281/zenodo.3963376.

\begin{figure}[htp]
\centerline{\includegraphics[width=9cm]{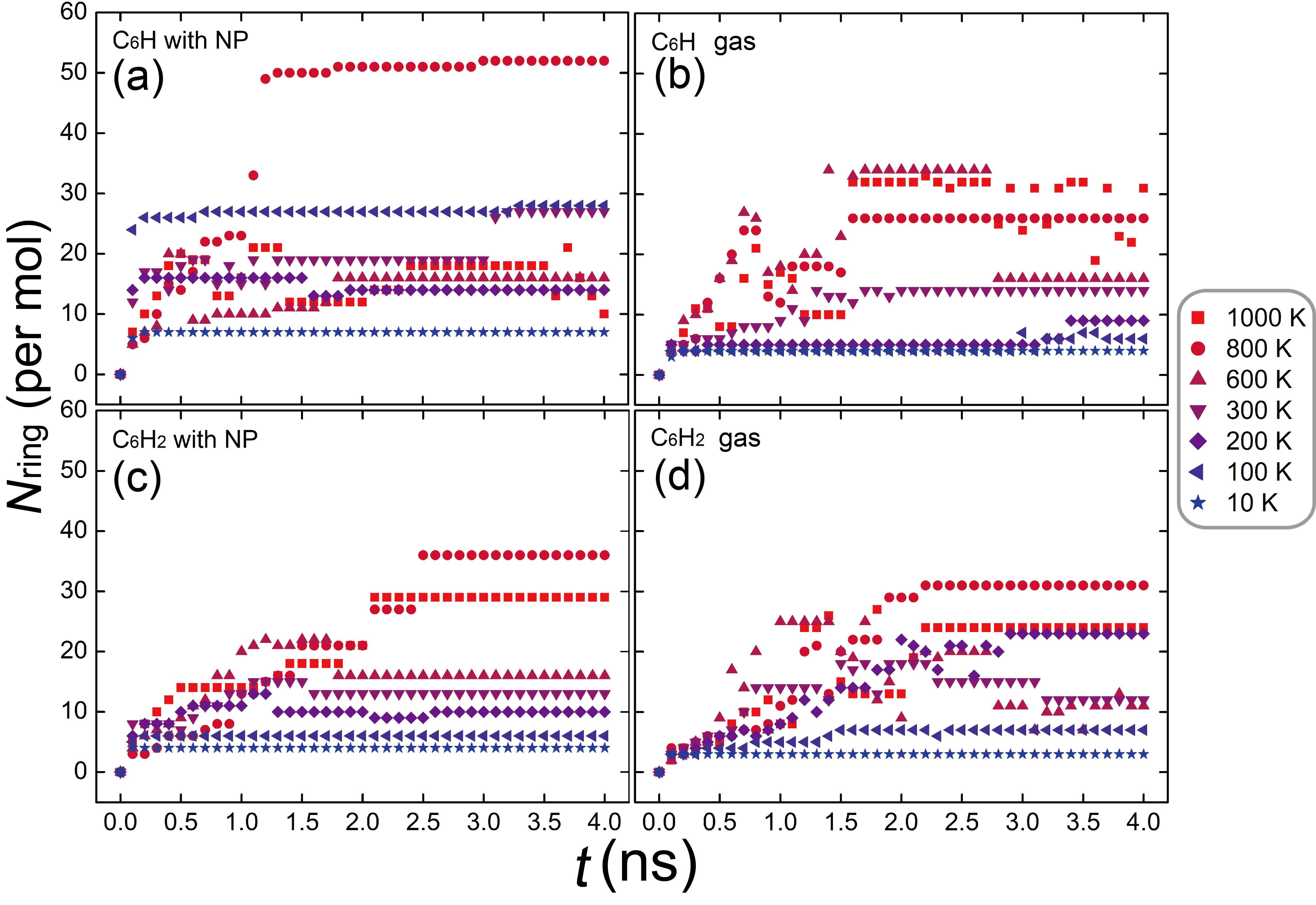}}
\caption{\label{F2}
Number of carbon rings $N_{ring}$ in the formed PAH molecules vs the simulation time $t$ for C$_{6}$H and C$_{6}$H$_{2}$ at different temperatures on the NP surface (a,c) or in the gas phase (b,d). The per-molecule value of $N_{ring}$ is averaged over all formed molecules in a simulation cell.}
\end{figure}

To measure the size of the complex structures, we compute the number of carbon rings $N_{ring}$ in each formed PAH molecule. Figure \ref{F2} shows the time-evolution of the averaged per-molecule value of $N_{ring}$ for C$_{6}$H (top panels) or C$_{6}$H$_{2}$ (bottom panels) at various temperatures. It is seen that the PAH size roughly increases with increasing temperature. This trend is notably kept in the cases of C$_{6}$H in the gas phase (panel b) and of C$_{6}$H$_{2}$ on the NP surface (panel c) at up to $800$ K. Figure \ref{F2} also shows a difference in the reactivity due to different levels of dehydrogenation, as the PAHs formed from C$_{6}$H (upper panels) are in general larger than those formed from C$_{6}$H$_{2}$ (lower panels). Moreover, the average size of the PAHs formed on the NP (left panels) seems to be slightly larger than that of the ones formed in the gas phase (right panels) at most of the temperatures studied here.

At low temperature, the nucleation of the PAH is hampered by slow diffusion and low collision rate, which result in saturated curves of $N_{ring}$ (horizontal lines in Figure \ref{F2}). In contrast, the size of formed molecules varies frequently at high temperatures, in particular at $1000$ K. This signifies the breaking or the merging of carbon rings, and thus indicates a possible change in the formation mechanism. Among the studied temperatures, the optimized temperature for forming large and stable PAH structures is $800$ K, for both the cases in the gas phase and on the NP. Note that Marshall and Sadeghpour have also shown by MD simulations that the formation of chain- or fullerene-like molecules is faster at $100-500$ K on a graphene surface than in the gas phase, but it becomes slower above $1000$ K \citep{Marshall2016}.

\begin{figure}[htp]
\centerline{\includegraphics[width=9cm]{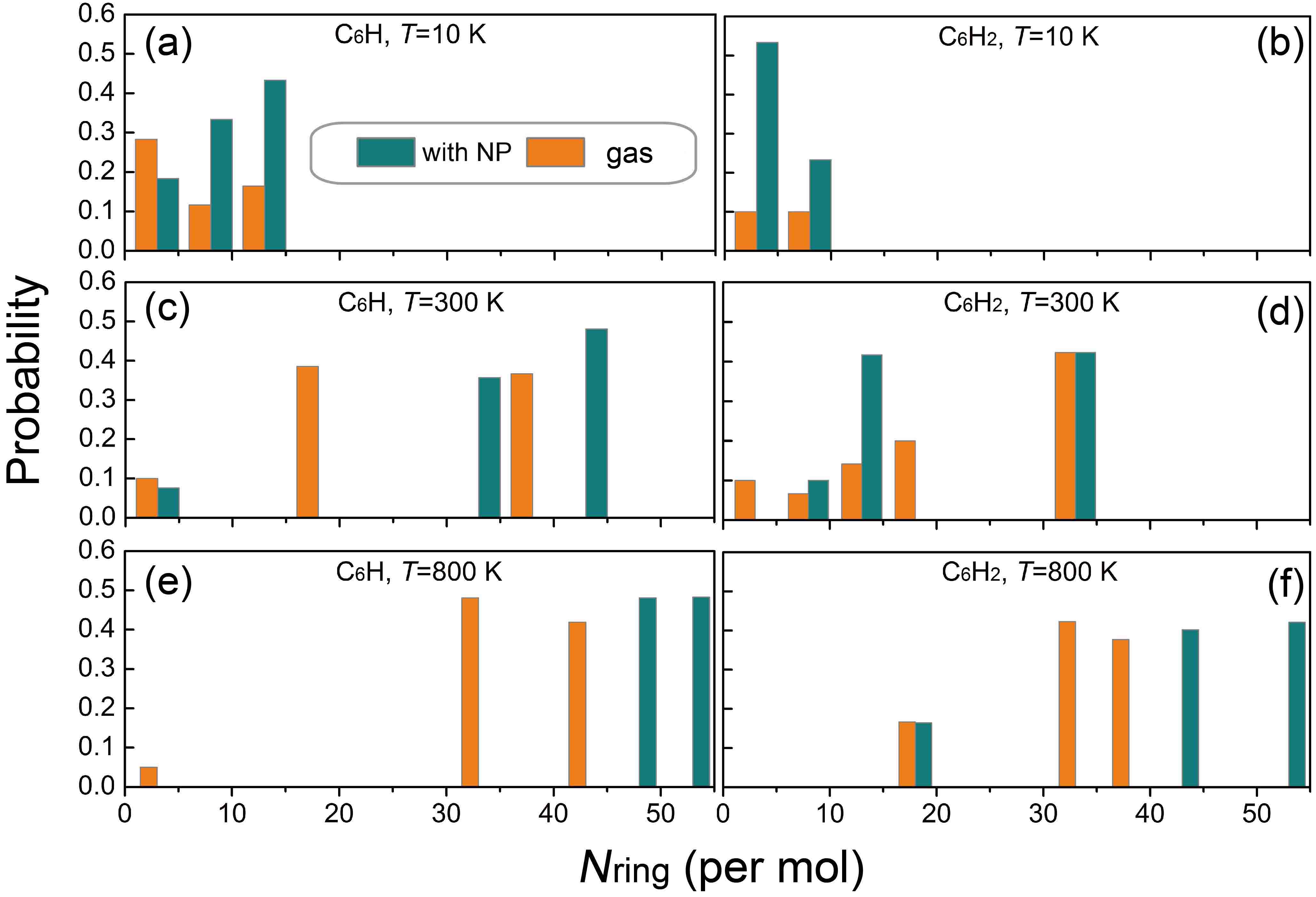}}
\caption{\label{F3}
Probability distribution of PAH molecules formed from C$_{6}$H (left panels) or C$_{6}$H$_{2}$ (right panels) at (a,b) $10$, (c,d) $300$, (e,f) $800$ K. The probability is calculated to be the ratio between the number of atoms in the formed molecule over the total number of atoms in the simulation cell.}
\end{figure}

The PAHs formed on the NP surface are found to be slightly larger than those formed in the gas phase. This effect can be more clearly seen within the plot of the probability distribution of the mean number of carbon rings in Figure \ref{F3}, which shows that large PAH molecules form more efficiently in presence of the NP. It can also be seen that the sum of probability is much smaller than $1.0$ for the low-temperature ($10$ K) case in the gas phase. This signifies that many of the benzene molecules do not participate the PAH synthesis due to their low mobility despite of the high molecular density.

\begin{figure}[htp]
\centerline{\includegraphics[width=9cm]{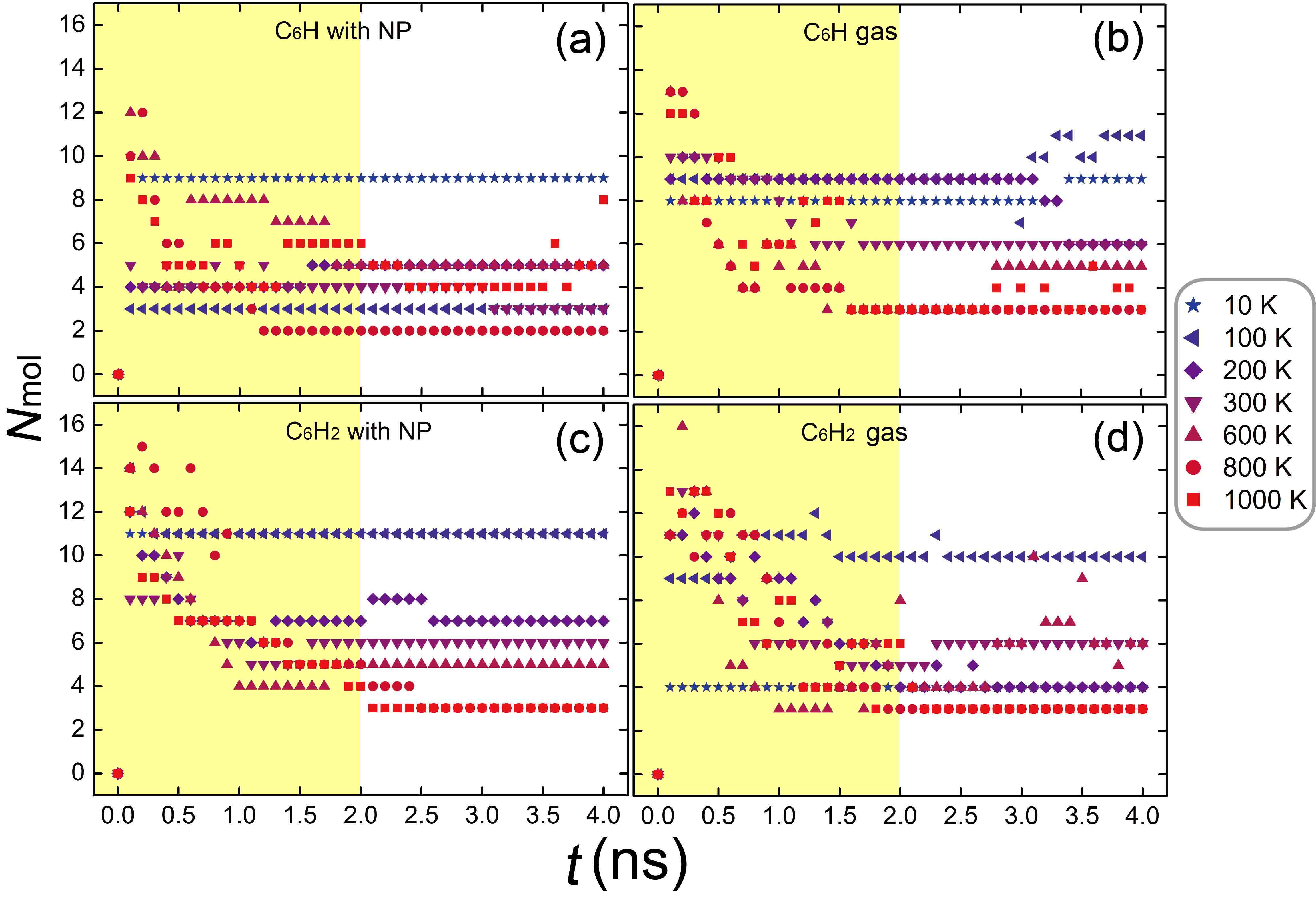}}
\caption{\label{F4}
Time-evolution of the number of the PAH molecules formed in the simulation cells.}
\end{figure}

One may wonder whether the PAH structures formed in $4.0$ ns will be the final products. To answer this question, we plot in Figure \ref{F4} the number of formed PAH molecules $N_{mol}$ as a function of the simulation time. It is seen that the evolution of $N_{mol}$ strongly depends on the applied temperature. For our simulation model, most of the reactions seem to take place within the first $2.0$ ns (highlighted by the colored zone). Further isomerization keeps taking place on a longer timescale with however much lower rate, particularly in the gas phase. At the optimized temperature of $800$ K that leads to large PAHs, the reaction seems to be possibly finalized within $2.5$ ns for all the four studied cases. Note that the discussion about the simulation time is ambiguous with respect to the real astronomical condition, since the simulations were substantially accelerated by using the extreme molecular density.

\subsection{Morphology analysis}

\begin{figure}[htp]
\centerline{\includegraphics[width=9cm]{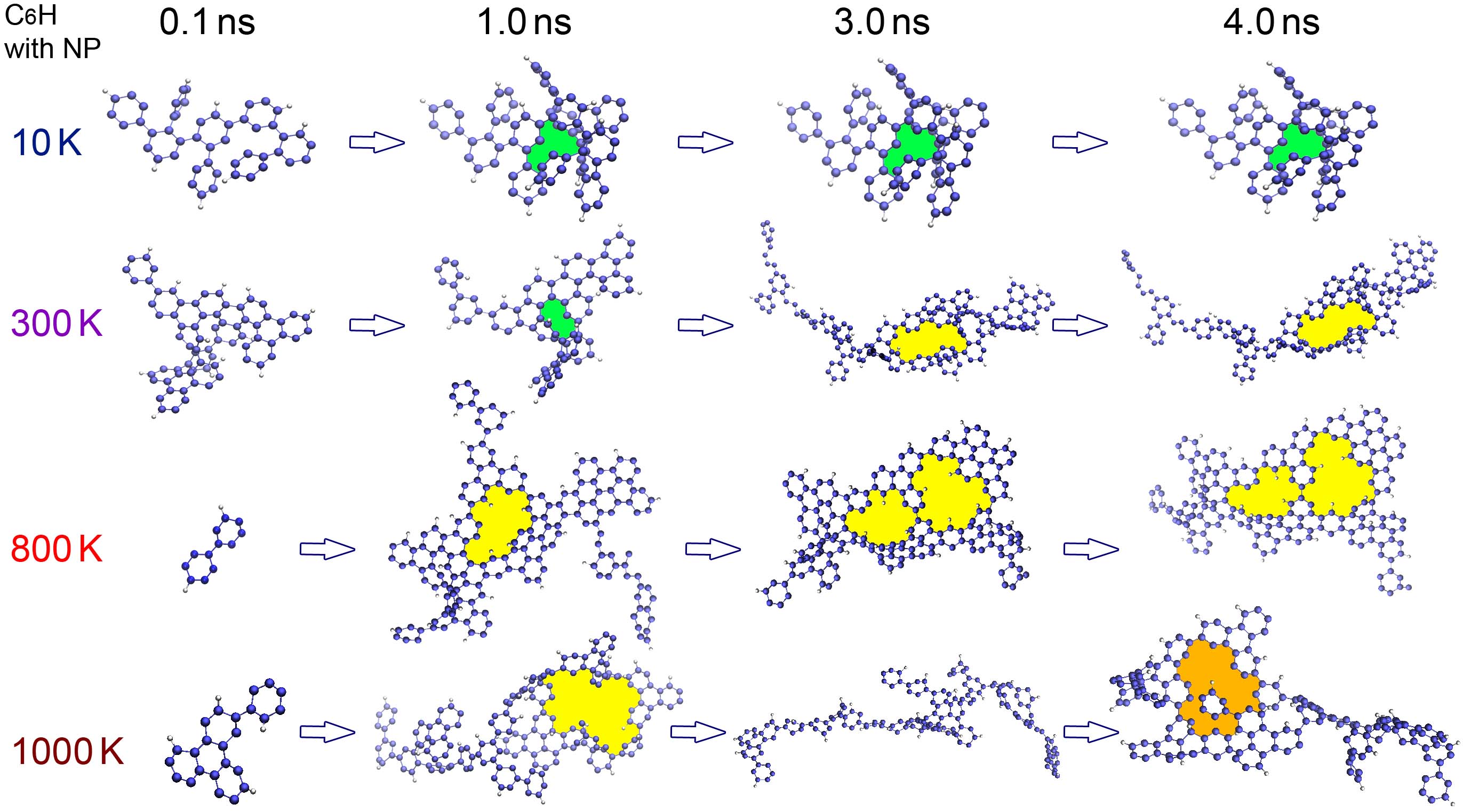}}
\caption{\label{F5}
Time-evolution of the morphology of PAH molecules formed from C$_{6}$H on the NP (not shown) at different temperatures.}
\end{figure}

\begin{figure}[htp]
\centerline{\includegraphics[width=9cm]{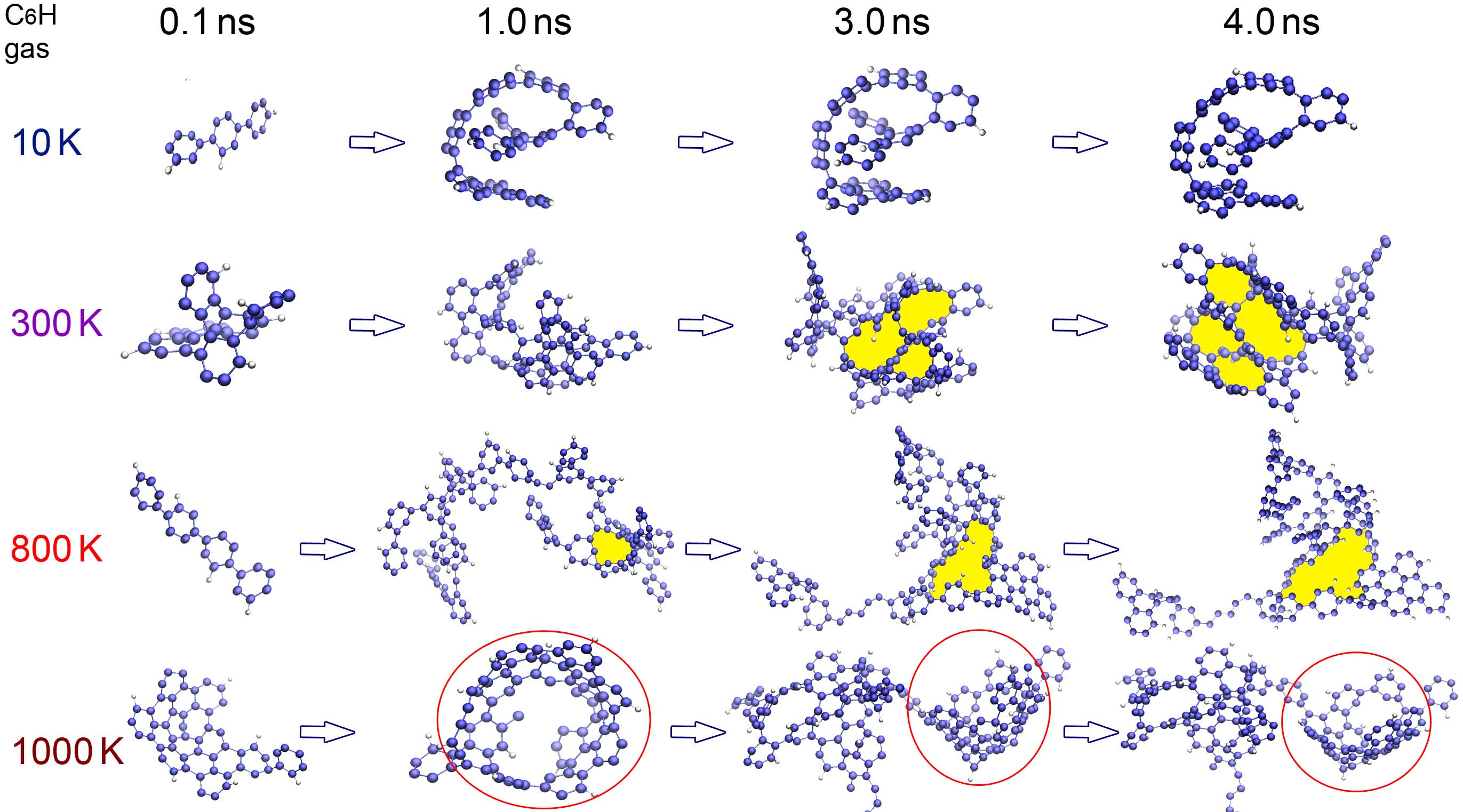}}
\caption{\label{F6}
Time-evolution of the morphology of PAH molecules formed from C$_{6}$H in gas phase at different temperatures.}
\end{figure}

In this subsection, we investigate the temperature effect on the PAH formation by analyzing the detailed morphology of the formed molecular structures. Figure \ref{F5} shows the growth of the PAHs formed on the NP surface over time from C$_{6}$H at four different temperatures. In keeping with the statistics shown in the previous section, the molecular structures formed at low temperature do not significantly change in the nanosecond time scale, but can quickly vary at high temperatures. The formed large carbon rings are highlighted by the colored patterns in Figure \ref{F5}. These rings {chaining more than $6$ C atoms efficiently form at high temperatures, and} maintain the connectivity of the typical $sp^{2}$ lattice in PAHs. They are unlikely to be stable since their sizes change over time. For instance at $1000$ K, a large C-ring is formed in $1.0$ ns by the collisions between small molecules, it then breaks in the following $2.0$ ns, before a new large C-ring is formed in $4.0$ ns by connecting suspended carbon chains.    

\begin{figure}[htp]
\centerline{\includegraphics[width=9cm]{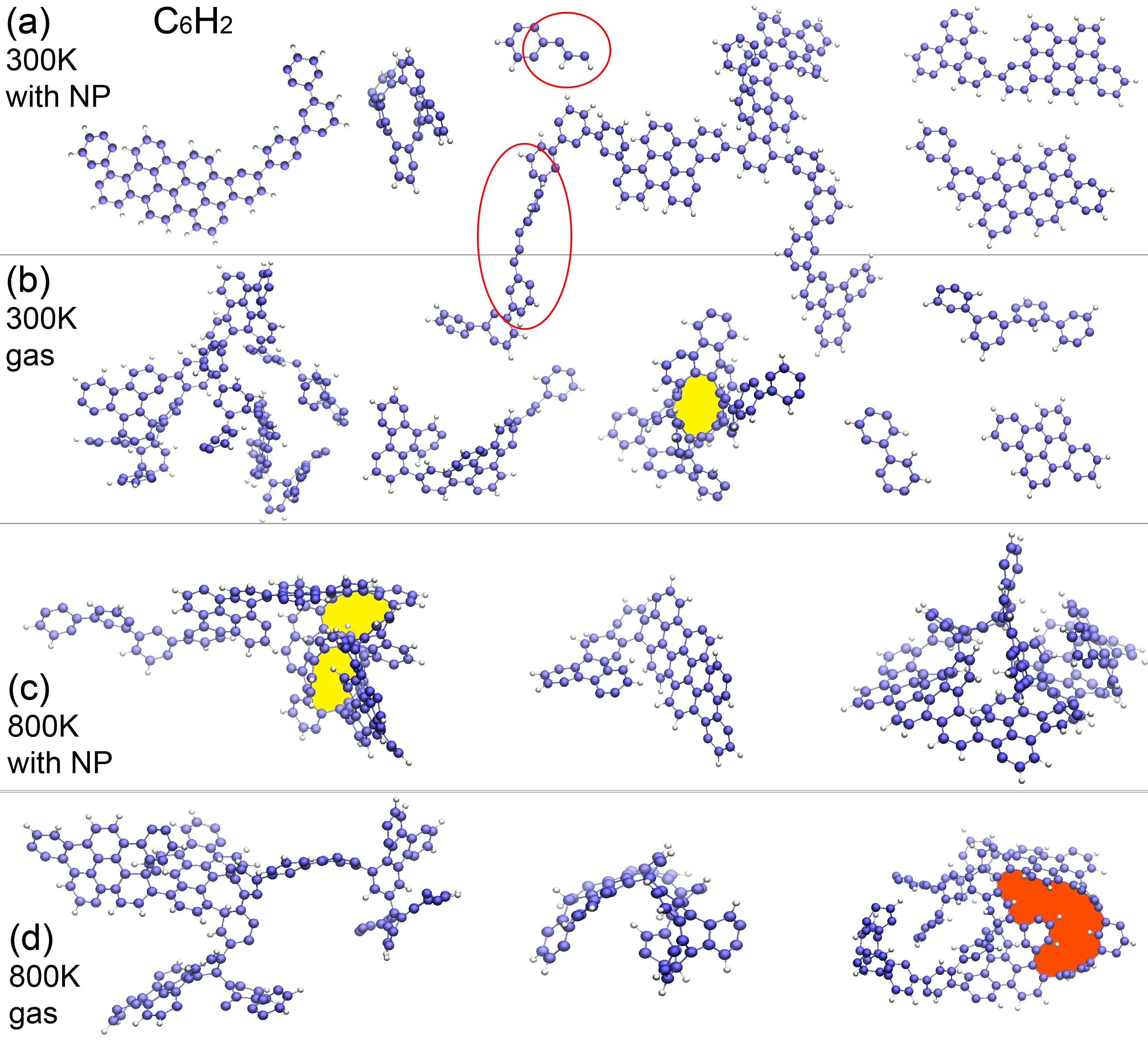}}
\caption{\label{F7}
Morphology of PAH molecules formed from $60$ C$_{6}$H$_{2}$ molecules at $300$ and $800$ K in $4.0$ ns.}
\end{figure}

It has long been recognized (by the so-called Mermin-Wagner theorem) that two-dimensional (2D) structures embedded in a 3D space have a tendency to be crumpled, since long-wave length fluctuations destroy the long-range order of 2D crystals \citep{Nelson2004}. It is seen in Figure \ref{F5} that most of the molecules formed on the NP are flatter than those formed in the gas phase, thanks to the supporting substrate. In contrast, the molecules formed in the gas phase are mostly curved, as shown in Figure \ref{F6}. This observation, together with aforementioned large PAHs formed on the NP surface (Figure \ref{F3}), may point to a possible layer-by-layer formation mechanism of fullerene and carbon onions, which have been reported to co-exist with PAHs in ISM \citep{Cami2010,Garcia-Hernandez2010,Chhowalla2003,Zhang2011}. 

An $e$-shaped molecule is formed at $10$ K by rolling up a flat PAH. We have performed structural optimization for this particular molecule by minimizing its potential energy, and found this curled structure is quite stable at a local potential minimum. Similar to the case on the NP, large carbon rings can also form in the gas phase as highlighted by the colored patterns, in particular at high temperature. Interestingly, a tubular structure is observed to form at $1000$ K in $1.0$ ns, as shown in the circles in Figure \ref{F6}. This tube-like molecule transforms into a cap-like structure in $3.0$ ns. This is in line with the simulations of Sadeghpour et al., in which curved cage-like carbon nanostructures were observed to form \citep{Marshall2016,Patra2014}.    

\begin{figure}[htp]
\centerline{\includegraphics[width=9cm]{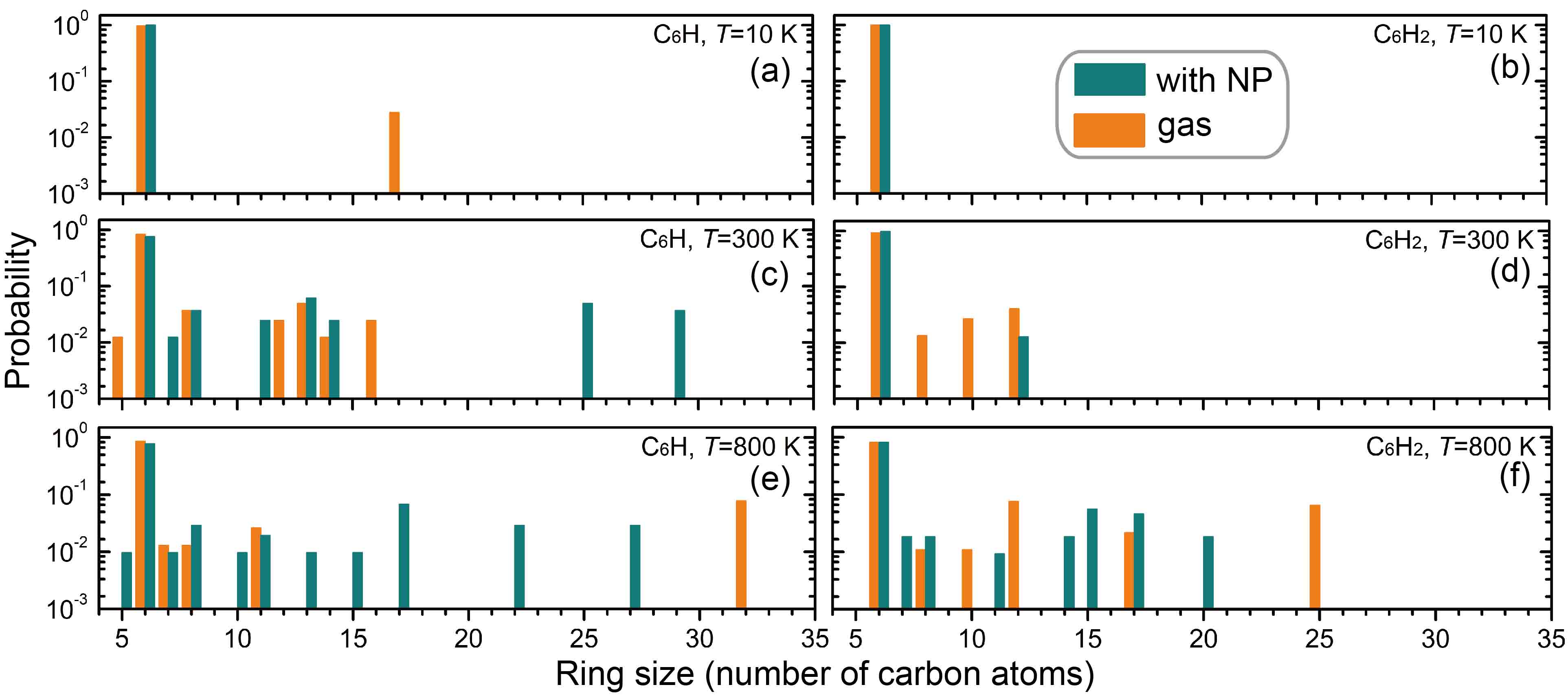}}
\caption{\label{F8}
Probability distribution of the number of carbon atoms in the rings of the PAH molecules formed from C$_{6}$H (left panels) or C$_{6}$H$_{2}$ (right panels) at (a,b) $10$, (c,d) $300$, (e,f) $800$ K. The probability is calculated as the ratio between the numbers of rings of a given size over the total number of rings of all formed molecules in a simulation cell.}
\end{figure}

Figure \ref{F7} shows the structures of molecules formed from C$_{6}$H$_{2}$ in $4.0$ ns. Similar to the case of C$_{6}$H, it can be seen that the PAH molecules formed in the gas phase are in general more crumpled than those formed on the NP, except for the small ones. We observe large carbon rings as well as mono-atomic chains (as shown in the red circles). The formation of these chains implies the breaking of carbon rings in the source molecules. This signifies that the dehydrogenation may induce structural instability of the carbon rings in certain conditions. A very large C-ring chaining $27$ carbon atoms (highlighted by the red pattern) is observed at $800$ K in the gas phase. Many of the sub-structures of these formed molecules are similar to those of the PAHs collected in the NASA AMES PAH IR spectroscopic database \citep{Bauschlicher2018}. We note that the molecular structures shown in this section only include a small portion of the PAH molecules ``formed'' in our simulations. A complete set of the structures of all new PAH molecules formed under different conditions are provided in supplemental materials \citep{SupplMater}. {Moreover}, it is also noted that these structures will make more sense for reaction in hydrogen-poor environment in ISM, since counterbalancing processes such as H-reaction were not taken into account in our simulations.

Figure \ref{F8} provides statistics on the ring size of the PAHs formed in $4.0$ ns. It is seen that the probability for producing large carbon rings increases with increasing temperatures. This probability is in general higher for C$_{6}$H than for C$_{6}$H$_{2}$. Most of these rings contain less than $17$ carbon atoms, including a number of pentagonal carbon (C-5) rings formed (e.g. from C$_{6}$H). Remarkably, a very large ring containing $32$ carbon atoms is formed at $800$ K from C$_{6}$H in the gas phase. We note that octagonal carbon (C-8) rings present in all the cases at $300$ and $800$ K for both C$_{6}$H and C$_{6}$H$_{2}$ in the gas phase and on the NP.

\subsection{DFT-calculated reaction pathways}

\begin{figure}[htp]
\centerline{\includegraphics[width=8cm]{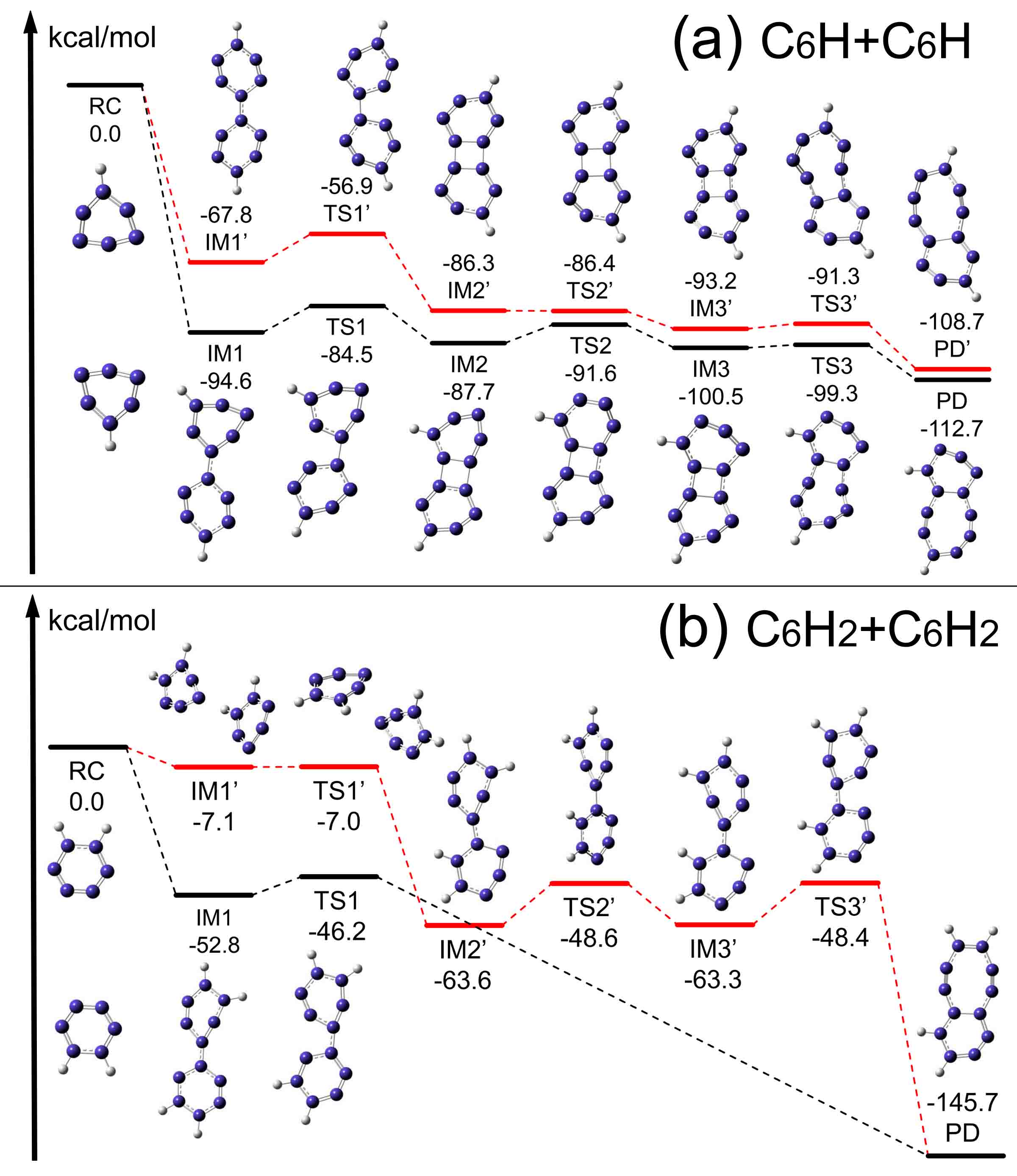}}
\caption{\label{F9}
DFT-computed free-energy diagram for the reaction between a pair of molecules of (a) C$_{6}$H$_{2}$ or (b) C$_{6}$H. The most energy-favorable reaction mechanism is highlighted in black color, while the second favorable one is in gray. All energy values are the Gibbs free energies at $15$ K in the unit of kcal/mol. Details on the optimized structures and their atomic coordinates are provided {in supplemental materials \citep{SupplMater}.}}
\end{figure}

To complement our MD simulations, first-principles DFT calculations are carried out to investigate the detailed chemical reaction pathway between individual pairs of molecules in single collision events at the atomistic level. The resulting molecular structures in the collision are optimized using the M06 functional \citep{Zhao2008} with 6-31+g(d, p) basis sets as implemented in Gaussian 16 B.01 \citep{Frisch2016}. Frequency calculations are also performed to confirm the intermediate (IM) or transition (TS) states). In addition, intrinsic reaction coordinate (IRC) calculations are carried out to make sure that TS actually connects the reasonable reactant (RC) and product (PD). Multiple (typically $8-10$) possible initial collision positions are considered for each pair of molecules, leading to multiple reaction pathways, among which the most energetically-favorable ones are selected and shown by the red lines in Figure \ref{F9}.

All of these reactions are exothermic and roughly barrierless. The energy released in the first step seems to be large enough to overcome all the subsequent energy barriers if the radiation loss is considerable. This indicates that even at temperatures as low as $10-20$ K in cold molecular clouds, these reactions can always occur due to the high chemical reactivity of unsaturated molecules. This is in agreement with \cite{Parker2015}, who have studied the formation of nitrogen-substituted PAHs. Furthermore, the large energy difference between the RC and the PD states implies there will be a high enthalpy released from chemical reaction to the kinetic energy of the formed molecule. In an interstellar environment with very low molecular density, this would suggest that the chemical reaction between PAHs could induce strong infrared emission signals, including even those from unstable intermediate or transitional molecular structures.

DFT calculations confirm the formation of PAHs containing stable C-8 rings through a serial of quasi-barrierless processes. It is found that the reaction between two C$_{6}$H$_{5}$ is the simplest, which needs direct collision between the reaction sites and it releases an amount of energy as large as $160.1$ kcal/mol. Figures \ref{F9} (a-b) show that the reactions between C$_{6}$H$_{2}$ or C$_{6}$H lead to formation of octagonal carbon rings (C-8). The energy barrier is $10.9$ kcal/mol for two C$_{6}$H and $6.6$ kcal/mol for two C$_{6}$H$_{2}$. When two radical monomers get close to each other, a covalent bond would be formed with releasing amounts of heat (-67.8 kcal/mol for C$_{6}$H+C$_{6}$H and -52.8 kcal/mol for C$_{6}$H$_{2}$+C$_{6}$H$_{2}$). The highest energy barriers along the most energetically-favorable reaction path are $10.9$ and $6.6$ kcal/mol when considering two C$_{6}$H and two C$_{6}$H$_{2}$, respectively.

\section{Conclusion} \label{sec:concl}

MD simulations suggest that the reaction between dehydrogenated benzene molecules can lead to the formation of complex PAH structures. The size of the PAH molecules roughly increases with increasing temperature in the gas phase or on the NP. An optimized temperature at the order of $800$ K is revealed in favor of forming large-size PAHs. The PAH size is also positively correlated with the level of dehydrogenation. Morphology analysis reveals the formation of large rings that contain more than $6$ carbon atoms. The probability for producing the large rings increases with increasing temperatures, and is in general higher for C$_{6}$H than for C$_{6}$H$_{2}$. DFT calculations identify the most possible reaction pathways, and quantitatively confirm the difference in the reactivity between C$_{6}$H and C$_{6}$H$_{2}$ as well as the formation of stable C-8 rings. We also observe the formation of (unstable) mono-atomic chains, as well as tubular and cap-like nanostructures. There is a general trend to form layered nanostructures of relatively large size on the NP surface, which points to a possible layer-by-layer formation mechanism for interstellar fullerene and carbon onions.\\


\end{document}